\documentclass[%
onecolumn,%
oneside,%
floats,%
aps,%
prd,%
nobibnotes,%
nofootinbib,%
amsmath,%
amssymb,%
amsfonts,%
amscd,%
superscriptaddress,%
eqsecnum%
]{revtex4-1}

\usepackage[utf8]{inputenc}
\usepackage{graphicx,array,dcolumn}
\usepackage[paperwidth=210mm,paperheight=297mm,centering,hmargin=1.8cm,vmargin=2.5cm]{geometry}
\usepackage{url}
\def\apj{ApJ}

\def\apjs{ApJS}

\def\aap{A\&A}

\def\azh{Astron. Zh.}
\def\jcap{J. Cosmol. Astropart. Phys.}

\def\mnras{MNRAS}

\def\nat{Nature}

\def\araa{ARA\&A}

\def\prd{Phys. Rev. D.}

\def\physrep{Phys.Rep.}

\def\ssr{Space Sci. Rev.}

\def\beq#1{\begin{equation}\label{#1}}
\def\eeq{\end{equation}}
\def\beqa#1{\begin{eqnarray}\label{#1}}
\def\eeqa{\end{eqnarray}}

\def\myfrac#1#2{\left(\frac{#1}{#2}\right)}

\def\comment#1{\relax}

\newcommand{\bi}{\bibitem}
\newcommand{\be}{\begin{equation}}
\newcommand{\ee}{\end{equation}}

\def\be{\begin{eqnarray}}
\def\ee{\end{eqnarray}}

\def\-g{\sqrt{-g}}

\renewcommand\rho{\varrho}
\renewcommand\tilde{\widetilde}

\begin{document}

\title{Antimatter and antistars in the universe and in the Galaxy}
\author{S.I. Blinnikov}
\email{Sergei.Blinnikov@itep.ru}
\affiliation{Novosibirsk State University, Novosibirsk, 630090, Russia}
\affiliation{ITEP, Bol. Cheremushkinsaya ul., 25, 113259 Moscow, Russia}
\affiliation{Kavli IPMU (WPI), Tokyo University, Kashiwa, Chiba, Japan}

\author{A.D. Dolgov}
\email{dolgov@fe.infn.it}
\affiliation{Novosibirsk State University, Novosibirsk, 630090, Russia}
\affiliation{ITEP, Bol. Cheremushkinsaya ul., 25, 113259 Moscow, Russia}
\affiliation{Dipartimento di Fisica e Scienze della Terra, Universit\`a degli Studi di Ferrara\\
Polo Scientifico e Tecnologico - Edificio C, Via Saragat 1, 44122 Ferrara, Italy}

\author{K. A. Postnov}
\email{pk@sai.msu.ru}
\affiliation{Moscow M.V. Lomonosov State University, Sternberg Astronomical Institute}

\begin{abstract}
We consider consequences of hypothetical existence of baryo-dense
stars created in the very early universe within an extension of Affleck-Dine scenario
of baryogenesis.
New constraints on the possible number of compact antimatter objects are derived.
The contemporary observational data do not exclude significant  amount of antimatter in the Galaxy (and in
other galaxies) in the form of  the  baryo-dense low-massive
stars.
\end{abstract}

\maketitle

\section{Introduction}
\label{sec:introduction}

Despite almost identical properties of particles and antiparticles, all matter observed in our neighborhood
consists only of particles, i.e. of protons, neutrons and electrons. A small fraction of antiprotons in cosmic rays,
about $10^{-4}$ with respect to protons,
most probably can be explained by their secondary origin.
Predominance of matter over antimatter was beautifully
explained by Sakharov~\cite{sakharov,*Sakharov1967JETPL}
as dynamically generated in the early universe due to breaking of C and CP
invariance, non-conservation of the baryonic number, and violation of thermal equilibrium.

On the other hand, there are plenty of theoretical possibilities leading to abundant creation of antimatter in the universe.
For example, if CP-invariance is broken spontaneously~\cite{1974PhR.....9..143L},
the universe would be equally populated by matter
and antimatter. However, in this  case the nearest antimatter domain should be practically at the cosmological
horizon, at a few gigaparsec distance~\cite{1998ApJ...495..539C}. 
Still less pessimistic scenarios are possible and independently of the
theory, a search for real (not secondary produced)
cosmic antimatter should be done and is being performed by several
detectors, including BESS~\cite{2008AdSpR..42..450S} 
PAMELA~\cite{2008JPhCS.110f2002B}, 
and AMS~\cite{1999PhLB..461..387A} 
More detectors are in project.

An unambiguous proof of existence of the primordial antimatter
would be an observation of sufficiently heavy antinuclei, starting from $^4$He.
According to theoretical estimates~\cite{2005PhRvD..71h3013D}
antideuterium could be created in the
energetic cosmic ray reactions of ${\bar p\, p}$ or ${\bar p}$~He collisions with a flux of
$\sim 10^{-7}$ m$^{2}$/ s$^{-1}$ /sr/(GeV/n),
i.e. five orders of magnitude lower than the observed flux of antiprotons.
The fluxes of the secondary-produced $\overline{^3\mbox{He}}$ and $\overline{^4\mbox{He}}$ are predicted to be
much smaller, respectively four and eight orders of magnitude below that of antideuterium. 
Experimental search of antinuclei production is performed at LHC by ALICE Collaboration. The results can be found in \cite{2013JPhCS.455a2007M}
and are reported at a seminar by
 A.P.~Kalweit \cite{Kalweit2014}.
Though the production rate looks significant, with the suppression factor about 1/300 per each extra (anti)nucleon added to
a nuclei, such events are quite rare in cosmology and their contribution to the total cosmological production is very small.

As for the astronomical observations, at the present time there are only upper bounds on the flux of cosmic
antihelium-4~\cite{2008AdSpR..42..450S}: 
$\overline{\mbox{He}} / \mbox{He} < 3\times 10^{-7}$.
In the nearest future this bound is expected to be improved down to
 $\overline{\mbox{He}} / \mbox{He} < 3\times 10^{-8}$~\cite{2008JPhCS.110f2002B} 
 and  ${\overline{\mbox{He}} / \mbox{He} < 10^{-9}}$~\cite{1999PhLB..461..387A}. 

There is another direction for search of cosmic antimatter through analysis of cosmic electromagnetic radiation, in particular
of $\sim 100$ MeV photons from $\bar p p$-annihilation and of the 0.511 MeV-line from $e^+e^-$ annihilation at low energies.
According to these data, the bounds on the fraction of antimatter in several galaxies, in particular on the amount of antistars, is
generally below $10^{-6}$ of the total amount of matter there.
The absence of excessive gamma radiation allows one to conclude that the
nearest {anti-galaxy} could not be closer than {$\sim$10 Mpc}~\cite{1976ARA&A..14..339S}. 
The mass fraction of antimatter in two
observed colliding galaxies in the Bullet cluster cannot be larger than $10^{-6}$~\cite{2008JCAP...10..001S}. 
As for our Galaxy, it is shown in ref.~\cite{2014HyInt.tmp...29V}, 
that the amount of antistars is bounded
by {${N_{\bar *} / N_{*} <  4\cdot 10^{-5} }$} within 150~pc from the Sun.

The quoted limits are valid if the antimatter objects are of the same kind as those made of the ordinary matter. However, they
may be essentially different, as argued in refs~\cite{ad-silk,ad-mk-nk},
where an efficient mechanism of the cosmological antimatter
creation was suggested and studied. According to this model, which is discussed below to make the paper self-contained,
the antistars could be formed in the very early universe as compact  objects, which may be predominantly dead now. They are not
concentrated in galaxies but distributed in a larger volume, in the halo, and have larger velocities than the normal stars.
In this sense, they are similar to cold dark
matter particles. In such a case the restrictive limits derived for the "normal" antistars are not applicable and such new type antimatter
objects may abundantly populate the Galaxy. Phenomenology of such antimatter objects is discussed in ref.~\cite{bambi-ad}.
As antistars may be abundant in the Galaxy, it is of interest
to consider the possibility of their registration by terrestrial detectors which
at the present time are sensitive
only to relatively close objects. It was shown in ref.~\cite{2012JETPL..96..421D} 
that the width of different atomic lines are
different for atoms and antiatoms, but the effect is too weak to be observed in the
foreseeable future. More promising for a registration  of antistars  in the Galaxy
are measurements of polarization of some lines of the stellar electromagnetic radiation, especially in the nuclear transitions
or of neutrino versus antineutrino fluxes~\cite{2014JETPL..98..519D}. 
There is no chance to see anti-stars in distant galaxies except for a lucky case
of a star-antistar collision which would be energetic enough to be observed on earth.

In ref.~\cite{2014PhRvD..89b1301D} 
we applied the mechanism of the early star formation, which is used below for creation of antimatter,
for an explanation of existence of quasars/supermassive black holes,  gamma-ray bursters and supernovae at high redshifts,
as well as of stars in the Galaxy, which look  older than the universe.

Some other models of the antimatter creation can be found in reviews~\cite{2000PhRvD..62h3505K,2002hep.ph...10012K}. 
The current observational bounds
are also extensively reviewed in refs.~\cite{2014HyInt.tmp...29V,2012NJPh...14i5012C} 

The standard scenarios of baryogenesis deal with only one number, namely with the ratio of the baryonic number density to the density
of photons in the microwave background radiation. Their usual outcome is a homogeneous baryon asymmetry all over the universe. So
there is no way to discriminate between different models having only one number in  possession. In this connection the models of
baryogenesis which lead to noticeable isocurvature perturbations and especially domains with negative asymmetry, i.e. antimatter domains,
are of great interest and an  astronomical search of cosmic antimatter and a study of theoretical scenarios
of antimatter creation become
of primary importance in the  attempts to understand generation of the cosmological matter-antimatter asymmetry.

The paper is organized as follows. In sec.~\ref{sec:ADscenario},  to make the presentation self-contained,
we discuss  the main features of the model leading to an efficient production of antimatter
objects (antistars, dense clouds of antimatter) which may "live" in the Galaxy in a significant amount.
Section~\ref{sec:diffusive} is devoted to estimates of possible abundance of antimatter in diffusive
interstellar state.
We derive new constraints on the possible number of compact antimatter objects in sec.~\ref{sec:compact}.
Discussion and conclusions are presented in sec.~\ref{sec:disc}.


\section{A mechanism of star and antistar formation in the early universe}
\label{sec:ADscenario}

All scenarios of baryogenesis, but one, normally predict rather low values of the baryon asymmetry, so theoretical efforts
are aimed at increasing the prediction up to the largest possible number to obtain the observed magnitude of the baryon
asymmetry
\beq{beta-obs}
\beta_{obs} = n_B/n_\gamma \approx 6 \cdot 10^{-10}\,.
\eeq
The only exception is the model suggested by Affleck and Dine (AD)~\cite{affleck-dine} where the
cosmological baryon asymmetry is normally much larger
than the observed value, and efforts should be done in the opposite direction to diminish it down to~(\ref{beta-obs}).

The AD-scenario of baryogenesis is based on high-energy supersymmetric (SUSY) model which naturally possesses flat directions
(valleys) in the potential of scalar superpartner of baryons. Such a scalar field has nonzero baryonic number and could acquire a large
vacuum expectation value during inflation, if its mass is smaller than the Hubble parameter at  that period. After inflation was over,
the baryonic number density accumulated in the rotational motion of the Affleck-Dine field, $\chi$, i.e. in its time dependent phase,
is transformed into baryonic number of quarks through baryo-conserved decays of $\chi$.

The essential features of the potential of $\chi$ can be described by the following toy model expression:
\beq{U-of-chi}
U(\chi)  = \lambda \left(  2 | \chi |^4  +  \chi^4 +  \chi*^4 \right) +
(m_1^2 \chi^2 + h.c.) + m_2^2 |\chi |^2\,,
\eeq
where the mass parameter  $m_1$ may be complex, leading to C and CP violation if $\lambda \neq 0$; $m_2$ may be taken
equal to $2|m_1|$ to avoid non-zero vacuum expectation value of $\chi$. A positive $m_2^2$ may appear as a result of
some spontaneous symmetry breaking  at a later stage of the cosmological evolution.
It is analogous to the well known Higgs phenomenon,
The first two  terms in the potential have flat directions, along which the potential does not grow, but for complex $m_1$ the
the flat directions of the quadratic term do not coincide with quartic ones.
Because of one-loop radiative corrections the quartic terms can acquire logarithmic factors
of the type $\sim \chi^4 \ln \left( {|\chi|^2 }/{\sigma^2 }\right)$~\cite{1973PhRvD...7.1888C}.  

During inflation, when the Hubble parameter was much larger than $|m_1|$ and $|m_2 |$,
the amplitude of field $\chi$ can reach high magnitude and when inflation was over and the Hubble parameter became
smaller than the slope of the potential, $\chi$ started evolving down to  the origin and on the way it could acquire a large "angular momentum"
in the complex $[ \mbox{Re}\, \chi, \mbox{Im}\, \chi]$-plane due to misalignment of the quartic and quadratic flat directions. The angular momentum is
essentially the baryonic number accumulated by $\chi$:
\beq{n-B-of-chi}
n_B (\chi) = i [ \chi^* (\partial_t \chi - (\partial_t \chi^*) \chi] .
\eeq
If there is no phase difference between $m$ and $\lambda$ and a quadratic flat direction coincides with a quartic one, then there would
be no misalignment but nevertheless a non-zero angular momentum could be induced due to quantum fluctuations of $\chi$ in the direction
orthogonal to the valley. In this case baryonic density would be also created but with different signs in different space domains because of
the chaotic behavior  of  quantum  fluctuations. So on average the baryon asymmetry would be zero.

This simple AD-scenario was slightly modified in ref.~\cite{ad-silk} by an addition of the general renormalizable coupling of $\chi$  to inflaton
field $\Phi$:
\beq{U-of-Phi-chi}
\delta U(\chi,\Phi)  = g|\chi|^2 (\Phi -\Phi_1)^2\,,
\eeq
where $\Phi_1$ is supposed to be the value which $\Phi$ passed during inflation not too close to its end. It allows
the size of the bubbles with high baron asymmetry to become astronomically large.
This is the only tuning parameter of the model. An essential effect
created by the addition of $\delta U$ to the potential of $\chi$
is that the window to the flat directions is open only for a short time when $\Phi$ is close to $\Phi_1$, so
the probability to reach a high value for $\chi$  would be small. Correspondingly $\chi$ would be large only in a small fraction of space.
In this case cosmologically small but possibly astronomically large
bubbles with high ${\beta}$ could be created, occupying a small
part of the universe volume, while the rest of the universe would have normal
${ \beta \approx 6\cdot 10^{-10}}$, created by a small ${\chi}$ which occupied the bulk of space. Nevertheless the fraction of baryonic and
antibaryonic matter in this compact objects may exceed that of the observed baryons. In the simplest version of the model the amount of baryons
and antibaryons in high $\beta$ regions
would be equal, but in more general case their ratio is model-dependent and may be arbitrary.

The bubbles with high values of $\chi$ after B-conserving
decay of $\chi$ into fermions would form domains with a large baryonic number
density in the form of the usual quark/baryon matter.
The rest of the universe would have normal small baryon asymmetry.
Initially the density contrast between the regions with low and high
values of $\chi$ was zero or very small (isocurvature perturbations).
After formation of the domains with large ${\chi}$
the equation of state inside and outside the bubbles became different
because the matter inside the bubbles was more non-relativistic than the matter outside. This would create some initial density
contrast  inside and outside high-B bubbles. For this and the following (next paragraph) conclusion it is essential that the baryon
diffusion length was very short.

The second period of generation of the density contrast
inside and outside the bubbles, ${\delta\rho}$, took place
after the QCD phase transition at temperatures somewhat above 100 MeV,
when quarks formed non-relativistic protons. At this stage a whole family of compact stellar-like objects
with baryon number density much higher than the background baryon density had been formed. Depending upon the
relation between their mass and the corresponding Jeans mass they could be very early stars, progenitors of
Supernovae, or primordial black holes. The mass distribution of these objects has the log-normal form:
\beq{ dn-dM }
\frac{dN}{dM} = C_M \exp{[-\gamma \ln^2 (M/M_0)]}
\eeq
where ${C_M}$, ${\gamma}$, and ${M_0}$ are constant  model-dependent
parameters. The form of the spectrum is practically model-independent since it is determined by the exponential expansion.
We call these regions as baryo-dense (BD) ones and the stars formed in such regions can be called  BD-stars and
even BD-black holes.

In particular, primordial black holes with
masses from a few solar masses up to ${10^{6-7} M_\odot}$, or even more
(on the tail of the distribution), can be created.
Such superheavy black holes could be seeds of galaxy formation. It is easy to choose the parameters
of the model that there would be one superheavy black hole (BH) for any
existing large galaxy. This scenario offers a new mechanism of the early superheavy BH
(quasar) formation and ultracompact dwarf galaxies where the BH may have more than 10\% of the
mass of the whole galaxy \cite{Seth2014}.
At the present time there is no satisfactory
explanation for the early formation of superheavy BH and ultracompact dwarf galaxies with large BHs
in the framework of the conventional theories.
Moreover, in the  model considered here a natural
explanation is found for existence of high red-shift gamma-ray bursters, of early supernovae, of the metal-
enriched chemistry
in the vicinity of those early-formed objects, and of the stars which are formally older that the universe.
These problems are discussed in an earlier publication by two of us~\cite{2014PhRvD..89b1301D}.
In this work we dwell on phenomenology, observational
manifestations, and bounds on antistars, which came to us from the early universe and might populate the
galaxy.

Of course big bang nucleosynthesis (BBN) in the regions with high $\beta$ would be significantly different
from the standard one with
low $\beta$; much more heavy elements would be produced there. The calculations of the element abundances
created at BBN with
high $ \beta$ have been done in ref.~\cite{BBN-high-beta}, but unfortunately only with $\beta \ll 1$, though with $\beta\gg \beta_{obs}$.
It would be very interesting to extend such calculations up to $\beta \geq 1$.  One immediate effect is that the hydrogen to
helium ratio would significantly drop down because for large $\beta$ the neutron-proton freezing took place at higher
temperatures when $n/p$ ratio was close to one. So the stars formed after BBN in high-B bubbles would mostly consist of $^4$He
plus some metals which are normally absent in  first stars. Since in the simplest version of the scenario the baryo-dense objects
consist of matter and antimatter in roughly equal number,
an anomaly in elemental abundances somewhere in the Galaxy could be an indicator of antimatter there with 50\% probability.
On the other hand, the observed abundances of light elements created at BBN
would not be significantly different from the predictions of the standard theory with
low $\beta$, because the low-$\beta$ volume is much larger than that with high $\beta$ and the clouds with anomalous abundances are
quite rare.

As is well known, the spectrum of angular fluctuations of CMB measures $\beta$ quite  close to (\ref{beta-obs}). However, it does not
exclude BD-objects considered here because the anomalies in the baryonic number density occur at very low scales much shorter
than tens megaparsecs to which the CMB measurements are sensitive.

In what follows we put aside theory and consider all possible observational manifestations of antimatter in the Galaxy not restricting
ourselves to any particular forms of antimatter objects, using the principle ``everything which is not forbidden is allowed''.

\section{Constraints from diffusive anti-matter}
\label{sec:diffusive}

Let us  start with the simplest constraint on diffusive anti-matter in the universe
from the production of high-energy photons (see also \cite{1976ARA&A..14..339S, 2014HyInt.tmp...29V}).

The observed isotropic EGRET background \cite{1998ApJ...494..523S, *2004ApJ...613..956S}
in the $\sim 100$~MeV energy range can be approximately represented as
\beq{}
I (100\, \mbox{MeV}) \approx 1000[\mbox{eV/cm}^2\mbox{/s/ster}]\,.
\eeq
This corresponds to the energy density of 100 MeV photons:
\beq{}
\epsilon_{obs} (100 \mbox{MeV}) =\frac{4 \pi}{c}  I (100 \mbox{MeV})  \approx 4 \times 10^{-7} [\mbox{eV/cm}^3] .
\eeq
Baryons in stars amount to $\Omega_b^*\simeq 10^{-3}$ \cite{1998ApJ...503..518F, *2004ApJ...616..643F} (in units of the critical density in the Universe),
which corresponds to the energy density
\beq{}
\epsilon^*=\Omega_b^*\rho_{cr}c^2\sim 5.2[\mbox{eV/cm}^3]\,.
\eeq

For the sake of a simple estimate we assume that  that most of the stars, including the BD ones, are similar to the Sun,
so they lose about 50\% of their mass in due course of evolution.
Adopting an equal amount of BD antimatter stars, $\Omega_{\tilde b}= \Omega_b^*$, and
allowing for the maximum annihilation of matter and antimatter during the Hubble time,
we get a rough upper limit on the spatial mixing of diffuse matter/antimatter, $f< 3\times 10^{-8}$,
in order to respect the observed gamma-ray background.

Note that unlike the usual stars, the BD stars can initially be compact
and consist of anti-helium, so their evolutionary mass loss can be  much smaller
than for the ordinary stars (see \cite{IbenTutukov1985,Nelemans_ea2001} and
the discussion below). Moreover, this mass loss took
place at high redshift, probably at $z>10$, and the flux
of the photons from matter and antimatter annihilation would be strongly reduced and shifted to smaller energies.
Therefore, it is necessary to consider another limiting case
where the primordial antimatter survived in the form of compact objects.

\section{Constraints from compact anti-matter objects}
\label{sec:compact}
\subsection{Annihilation at accretion}

A compact BD-star with mass $M$ passing through diffusive interstellar
or intergalactic medium with number density $n_0$ will
accrete baryonic matter. According to the well-known Bondi-Hoyle-Littleton formula,
the accretion rate is
\beq{BHL}
\dot M\simeq \myfrac{2GM}{v^2}^2 m_p n_0 v\approx 10^{11} [\hbox{g/s}]
\myfrac{M}{M_\odot}^2 \myfrac{n_0}{1\,\hbox{cm}^{-3}} \myfrac{v}{10 \,\hbox{km/s}}^{-3}\,.
\eeq
This accretion rate exactly corresponds to the widely used formula (23) from ref.
\citep{1976ARA&A..14..339S} for photonic luminosity due to annihilation of accreting matter.

As mentioned in Section \ref{sec:introduction},
compact BD stars can be treated as cold dark matter particles. That is, they should
have virial velocities in galaxy halos about $v_{\rm BD}\sim 500$~km~s$^{-1}$.
Unfortunately the galactic escape velocity {and} hence the virial velocity are not well known
\cite{1981ApJ...251...61C,1988PhRvD..37.2703G}.
{ For a recent review see ref.~\cite{2012MPLA...2730004G}.}
{ The} value $v_{\rm esc} =650 \, {\rm km \, s}^{-1}$
($90\%$ upper confidence limit) from ref,~\cite{1990ApJ...353..486L} is { usually taken}.
Ref.~\cite{2007MNRAS.379..755S} finds the updated escape speed in the range
$498 \, {\rm km \, s}^{-1} < v_{\rm esc} < 608 \, {\rm km \, s}^{-1}$
at $90\%$ confidence, with a median likelihood of $v_{\rm esc} =544 \,
{\rm km \, s}^{-1}$.

Therefore, the gas accretion rate onto a compact BD-star is
dramatically decreased, $\dot M_{\rm BD}\sim 10^6$~g/s if we take safe realistic values
$v_{\rm BD}\sim 300-500$~km~s$^{-1}$.
This implies that BD-stars comply with { the} constraints considered
in ref~\cite{2014HyInt.tmp...29V}.

At this low accretion luminosity ($\sim 10^{27}-10^{28}$~erg/s, which may be lower
than their intrinsic luminosity), BD-stars are very difficult to discover.
They may appear as rapidly moving cold objects, for example, in the forthcoming GAIA or WSRT observations.

\subsection{Binary BD stars}

Consider now a  possible case of a binary BD-star consisting of antimatter. In the process of evolution, the stars in the system
could coalesce due to gravitational wave emission to produce an explosive event like
ordinary type Ia supernova.
This explosion would inject around 1-2 solar masses of antimatter into the interstellar medium and
produce a supernova remnant (SNR).
In addition to ordinary thermal shock-wave emission (mostly in keV range),
in this case one should expect a   high flux of hard photons from $e^+e^-$ and proton-antiproton annihilation in the
interstellar medium.

In reality, the mean free-path of an (anti)baryon is determined by
the magnetic field which is inevitably generated behind the shock front
\cite{1990acr..book.....B,1987PhR...154....1B,1991SSRv...58..259J,2004MNRAS.353..550B,2005AstL...31..748B,
*2006ApJ...652.1246V,*2007PhyU...50..141B,*2012SSRv..166...71B}.
For an estimate, assume $B\equiv 10^{-5}\hbox{G}\times B_{-5}$. The Larmor radius
for a proton is
\beq{}
r_L=(v/c) m_pc^2/eB=10^{10}[\hbox{cm}] \,B_{-5}^{-1} v_9\,,
\eeq
where the shock velocity, $v_9$,  is normalized to 10000 km/s.

Let us consider a young anti-SNR with radius $R=10^{18}\hbox{cm}\, R_{18}$.
The mass inside the layer of active annihilation is equal to:
\beq{}
\Delta M =4\pi R^2 r_L m_p (4 n_0)\,,
\eeq
where $n_0\sim 1$ cm$^{-3}$ is the ambient ISM
number density. The annihilation time is $t_{\rm ann}=1/(n\sigma v)$, where
the annihilation cross-section is inversely proportional to the center of mass  velocity of the colliding proton-antiproton pair and is
given  by  $\sigma v \sim 10^{-15}$~cm$^3$/s. So the annihilation time is:
\be
t_{\rm ann} \approx 10^{15}[\hbox{s}] (n/\hbox{cm}^3)^{-1}
\label{t-ann}
\ee
In the case of  very slow relative velocity of protons and antiprotons, e.g. $v \sim 10^{-3}$, the cross-section
is enhanced approximately by an order of magnitude due to Coulomb attraction between $p$ and $\bar p$
(Sommerfeld enhancement).
Thus, the annihilation luminosity is
\beq{Lann}
L_{\rm ann}=\frac{\Delta Mc^2}{t_{\rm ann}} \approx
5 \cdot 10^{29} [\hbox{erg/s}]\, R_{18}^2  B_{-5}^{-1} (n/ \hbox{cm}^3)^2 .
\eeq
The absolute upper limit on the annihilation luminosity is
$\sim M_\odot c^2/t_{\rm ann} \approx 10^{39}$~erg/s, but, of course, the estimate
(\ref{Lann}) is much closer to reality.

There can be also binaries formed of a star and an antistar by the gravitational capture through
a three-body interaction with another star.
The probability of such binary formation is possibly much smaller than the probability of formation of
a binary consisting of two antistars which could be created
from the same cloud with high baryon density.
Still it is nonzero, and merging of antistar and star would lead
to a  spectacular explosion (hypernova?).

\subsection{Microlensing}

Compact BD-stars can be also found by the effect of
gravitational microlensing which may be caused by both visible and invisible stars.
These objects are called  now  {\sc macho}s for ``Massive Astrophysical
Compact Halo Object''.
This phenomenon was first discussed in relation with DM candidates made of the so-called Mirror Matter
by Berezhiani, Dolgov, and Mohapatra~\cite{Berezhiani1996}
and Blinnikov~\cite{Blinnikov1998}). 

\subsubsection{MACHO, EROS, AGAPE, MEGA, OGLE -- contradicting results}

MACHO group \cite{MACHO2000} has revealed 13 - 17 microlensing events in the Large Magellanic Cloud (LMC),
significantly higher number than { that} expected from { the} known stars but not enough to explain all dark
matter (DM) in the halo.
The fractional contribution of the objects which produced the lensing into the dark matter density
is usually denoted as $f$. Such objects have also got the name Machos.
MACHO group concluded that compact objects in the mass range
$ { 0.15M_\odot < M < 0.9M_\odot } $ have a fraction $f$ in galactic  halo
in the range ${ 0.08<f<0.50}$ (95\% CL).
So Bennett \cite{Bennett2005} has concluded (based on the results of MACHO group) that
Machos have been really found.

EROS (Exp{\'e}rience pour la Recherche d'Objets Sombres) collaboration  has placed only an upper
limit on the halo fraction, $f<0.2$ (95\% CL) for { the} objects in the specified above MACHO
mass range, while EROS-2 \cite{EROS2007} gives $ f<0.1$ for $10^{-6}M_\odot<M<1M_\odot$.

AGAPE collaboration \cite{AGAPE2008}, working on microlensing in M31 (Andromeda) galaxy, finds
the halo { Macho} fraction in the range $0.2<f<0.9$.
while MEGA group marginally conflicts with them with an upper limit $f<0.3$
\cite{MEGA2007}.

Detailed analysis of the controversial situation with the results of different groups is given in
ref.~\cite{Moniez2010}.
Newer results \cite{OGLE2013} for EROS-2 and OGLE
(Optical Gravitational Lensing Experiment)
in the direction of the Small Magellanic Cloud are:
 $ {f <0.1} $ obtained at 95\% confidence level for { Macho}s with the mass $ 10^{-2} M_\odot$
and $ {f <0.2} $ for { Macho}s with the mass $ 0.5 M_\odot $.

Recent data on  the other aspects  of the microlensing are discussed in ref.~\cite{Mao2012}.

It would be exciting if all DM were constituted by BD-stars (and BD black holes) with masses in still allowed
intervals,  but more detailed analysis of this possibility has to be done.

\subsubsection{Destruction of wide pairs of visible stars}

Paper \cite{YooCG2004}, which appeared in the series ``End of MACHO era (1974-2004)'',
asserts that wide pairs
of visible stars must be destroyed by invisible {Macho}s flying  near them.
The same effect may take place in the  case of BD-stars (which are visible, but weak).
In addition to the criticism of the paper~\cite{YooCG2004}, put forward in~\cite{EvansBel2007},
one can point out that it is necessary to consider not only
the process of destruction, but also a reverse process of creating pairs of visible stars from single individual
stars not { bound} previously by the mutual gravity.

The probability of microlensing \cite{Byalko1969,*Byalko1970,Pacz1986} is naturally measured by the
so-called optical depth $ \tau $.
Evans and Belokurov \cite{EvansBel2007} confirmed lower number of compact objects in the direction to LMC,
than obtained in the MACHO group, i.e. they got $\tau <0.36 \times 10^{ -7 } $ in agreement with EROS
results  \cite{EROS2007}.
Later, however, a paper of the same Cambridge group \cite{Quinn2009} was published where, on the basis of
studies of binary stars, arguments in favor of real existence of {Macho}s and against the pessimistic
conclusions of ref.~\cite{YooCG2004} were presented.

\subsection{Reionization and CMB} \label{s:reionization}

Energy influx from matter-antimatter annihilation at high redshift, $z>10$ could be a source of the cosmological reionization, for
which not enough energy is found in the standard model.  Another way around, we can obtain a bound on the amount of antimatter
annihilating at high redshifts, if no energy injection is observed. A simple constraint can be derived as follows. According to
eq.~(\ref{t-ann}) the annihilation time in the early universe is
\be
t_{\rm ann} (z)  \sim 0.5 \cdot 10^{23} [\rm{s}](1+z)^{-3}
\label{t-ann-of-z}
\ee
Here we took for the baryon number density the value $n = \Omega_b \rho_c /m_p \approx 2\cdot 10^{-7} (z+1)^3$.
An account of the Sommerfeld enhancement, which is effective at low velocities, $v<10^{-3}$, could enlarge the cross-section by an order of
magnitude and correspondingly diminish $t_{\rm ann}$ by the same factor.

At redshift $z=10$ the annihilation time is about $4\cdot 10^{19} $ seconds, which is much longer than the universe age at this redshift,
$t_U (z=10)  \approx 10^{16} $ s. So one may expect that about $10^{-3} -10^{-4}$ of all diffuse antimatter could annihilate, producing mostly
100 MeV photons, energetic electrons and positrons. In principle there would be more than enough energy to reionize the universe but more
detailed calculations of the rate of energy degradation down to atomic resonance is necessary, which is outside of the scope of the
present work.

The universe age became close to the annihilation time near the hydrogen recombination era at $z=1100$ or earlier. The annihilation
could distort the CMB frequency spectrum if the annihilation proceeded below $z=10^6$. However, the energy of the photons produced
by $p\bar p$ or even $e^-e^+$ annihilations is by far above the CMB energy at this epochs. Moreover, we believe that practically all
antimatter was confined inside BD stars and very little annihilation proceeded on the star surface.
However, e.g. a helium BD-star with the mass $M_{\rm He}\sim 2.5 M_\odot$, would evolve only about 500000 years, so
most of the mass-loss would occur shortly after recombination, when the universe age was comparable to the annihilation time
so the annihilation energy would interfere with the recombination dynamics, which may lead to some observable effects to be studied.

\subsection{Meteor observations}
\label{s:meteors}

As was mentioned above in Sec. II, (some of) the BD-stars may potentially have anomalous chemical abundances
due to different values of the baryon asymmetry parameter $\beta$ during the primordial nucleosynthesis. If $\beta$ is low, no
significant differences as compared to the standard BBN abundances are expected. Such 'metal-free' BD-stars
should be similar to the first Population III stars.
Not much dust is expected to be formed during evolution of these stars.
However, in domains with large $\beta$ the initial metal abundance can be higher.
These 'metal-rich' BD-stars in due course of the evolution can produce a certain
fraction of solids mostly in the form of dust particles.
These dust particles should move with virial
halo velocities $v\sim 500$~km/s and can be observed as  'anti-meteors'.
An anti-meteor with mass $m$ intruding the Earth atmosphere should produce
a prolific gamma-ray emission with a fluence of about $F_{\gamma}\sim 10 (m/1\mathrm{mg})$~erg/cm$^2$ on a
time-scale of the Earth atmosphere crossing
$\sim 0.1-1$~s. The non-detection of such bright flares by all-sky gamma-ray monitors
from the Earth atmosphere and from the Moon\footnote{{The brightest gamma-ray bursts observed by BATSE have
the fluence of $\sim 10^{-3}$~erg/cm$^2$ \cite{1999ApJS..122..465P}.}} led Fargion and Khlopov
\cite{2003APh....19..441F}
to infer the upper bound on the antimatter fraction in meteors to be
$f<10^{-8}-10^{-9}$, {which is derived by assuming the complete symmetry between matter and
antimatter.} (Note that short, millisecond, intensive
hard gamma-ray
flashes with energy $10^8-10^9$~ergs are observed from the atmosphere (the so-called
'terrestrial gamma-ray flashes', TGF)\cite{1994Sci...264.1313F} and are thought to be
associated with atmospheric electricity \cite{2012SSRv..173..133D}.)

An extragalactic meteor
intruding the Earth atmosphere with a velocity of at least 300 km/s was
reported in  ref.~\cite{2007AstBu..62..301A}. The spectrum of this faint meteor
was similar to the standard metal-rich chondrite. The authors \cite{2007AstBu..62..301A} concluded
that the space number density of such rapid dust particles, apparently of extragalactic origin,
in the vicinity of the Earth can be as high as $n_d\sim 4\times 10^{-26}$~cm$^{-3}$. The
(very model-dependent) estimate of the observed meteor mass is $m\sim 10^{-6}$~g.
The prolific
metal-rich extragalactic dust particles is quite enigmatic from the point of view of standard
stellar evolution.
If this event were an anti-meteor, the associated gamma-ray fluence, assuming a 500~km distance to the space
detector, would be about $10^{-2}$~erg/cm$^2$, which is several orders of magnitude higher than the
brightest BATSE flashes.

According to the standard belief the first stars ejected molecules and dust. To form larger pieces of matter (meteors)
such gas should be compressed e.g. by a shock wave from supernova explosion or by a collision with another
molecular cloud. As a result dust particles could be squeezed  forming larger stones or pieces of ice forming
protoplanetary or protostar clouds.  However, the BD stars not necessarily passed through such cycles but most probably
remained primary (PopIII) stars. In this case one should not expect large number of meteors from them in our neighborhood.

Thus in the context of BD-stars under scrutiny
the anti-matter restrictions derived in
\cite{2003APh....19..441F} from meteor observations cannot be directly applied. Indeed,
most of the cosmic anti-dust produced by mass-loss from 'metal-rich' BD-stars is
expected to
be in the form of small micron-size grains with a mass of $10^{-6}$~g and smaller, like
the ordinary interstellar dust. Estimates made in \cite{2003APh....19..441F}
show that these small anti-dust grains should completely annihilate when
moving through the interstellar gas, contributing to the diffusive gamma-ray background.
Guided by the analogy with the ordinary matter, the fraction of dust in the total stellar mass
should be smaller than $\sim 1\%$ \cite{2004ApJ...616..643F}, thus providing only minor contribution to the
observed gamma-ray background discussed above.

\section{Discussion and Conclusions}
\label{sec:disc}

We conclude that the contemporary observational data do not exclude significant  amount of antimatter in the
Galaxy (and in other
galaxies), especially in the form of  the  baryo-dense {low-massive}
stars created in the very early universe. The total mass of these antimatter
objects could be comparable with the total mass of the Galaxy. They would populate the galactic halo and might
make a noticeable
contribution to dark matter and, in particular, to Machos observed through microlensing. The BD stars should
have an unusual chemical
content because they were formed in the regions with very high baryon-to-photon ratio, where BBN proceeded
with more efficient synthesis
of heavy elements. Thus a star with chemical anomaly may present a good possibility to be an antimatter star.

As we have shown above, the diffusive gamma-ray background
may impose  stringent constraints on the BD-star fraction because of inevitable mass-loss
during evolution of the anti-helium stars. The precise amount of mass lost during evolution of an (anti)
helium star
depends upon its mass and metal abundance (see e.g. \cite{IbenTutukov1985} for more detail). For example, for
solar metal abundance if
$M_{\rm He}\leq 0.3 M_\odot$,  helium ignition into carbon is impossible,
and the star simply cools down to form an anti-helium white dwarf without mass-loss.   It is also known
that for $M_{\rm He}\leq 0.8 M_\odot$ no significant mass loss is expected, and the star is evolved to form a
hybrid CO-He white dwarf
\cite{IbenTutukov1985}. Helium stars with initial masses $0.8<M_{\rm He}/M_\odot\leq 2.2$ evolve to form a CO
white dwarf with a mass up to 1.2 $M_\odot$,
i.e. the mass-loss from helium stars in this mass interval increases from 0 to about 45\%. Cores of more
massive helium stars are expected to collapse to form a neutron star or black hole (for the most massive
stars). Whether the collapse into a black hole is accompanied with substantial mass loss is unclear.
The nuclear evolution of more massive helium stars occurs on the timescale
$\tau_{\rm He}\sim 10^{7.15}(M_{\rm He}/M_\odot)^{-3.7}$~yrs \cite{Nelemans_ea2001}. Therefore, for
interesting He-star masses
$M_{\rm He}\gtrsim 0.8 M_\odot$, most of the possible mass loss is expected to occur in the first  60 mln
years, i.e. at redshifts $z\gtrsim 43$ (for the standard cosmology), long before the formation of first
structures in the universe.
Assuming a homogeneous medium with the density
$n(z)=n_0(1+z)^3$, where $n_0=\rho_b/m_p\sim  2\cdot 10^{-7}$~cm$^{-3}$
is the present-day baryon number density, taking the annihilation cross section to be
$\sigma_{\rm ann}\sim 10^{-24}(c/v)$~cm$^2$ (without account for the  Sommerfeld low-velocity enhancement),
and using the thermal velocity $v_{\rm th}\sim 10^5 \sqrt{T/10^4K}$~cm/s in
adiabatically cooling ideal mono-atomic gas with $T(z)=3\times 10^3 [(1+z)/10^3]^2$~K, we can estimate the
annihilation time $t_{\rm ann}\sim 1/(n\sigma_{\rm ann} v)$ to be much longer than the Hubble time at the
corresponding $z$. This implies that most of the gas anti-matter
from BD stellar winds is likely to survive until the present time. Even at the present time, in dense
intercluster gas with baryon number density $n\sim 10^{-3}$~cm$^{-3}$,
such anti-particles moving with virial velocities of a few 1000~km/s can annihilate only very slowly.

The physics of BD-stars is quite poorly studied and may be very  much different from the usual
astrophysics because the initial states of such
stars quite often were different from the initial states of the usual stars. For example, BD-stars could be
formed in the state when the external
pressure was larger than the internal one. Moreover they start from already dense and hot state but not from
cold disperse gas cloud.
In particular, there could be BD-stars which are similar to the core of red giants but without external layers
and some other strange objects, see discussion in ref.~\cite{bambi-ad}.

Let us  note that allowing some fraction of the anti-matter to annihilate at high redshift may contribute to the hard radiation continuum
which is necessary to the secondary ionization of the Universe at $z>10$.

As is discussed in the previous papers, BD stars may have age comparable to the universe age, and due to an anomalous initial chemistry
they even might look older than the universe if their age is determined  by the usual methods under assumption of the standard
initial chemistry. This problem is under investigation. The observed high-redshift supernovae and gamma-ray bursts also nicely
fit into the frameworks of the model.

As a by-product, the model considered here explains the formation of superheavy black holes and suggests an inverted process
of the galaxy formation: first, a superheavy BH was born, which served as a seed for the subsequent collection of matter making galaxy.

\acknowledgements
We are grateful to A.V.Zasov for useful indications on DM literature and to M.S. Pshirkov and L.R. Yungelson for discussions.
SB and AD acknowledge the support of the grant of the Russian Federation government 11.G34.31.0047.
{The work of KP was supported by the Russian Science Foundation grant 14-12-00203.}


\bibliography{antistars.bib}
\end{document}